\documentclass[11pt,a4paper]{article}

\usepackage{jheppub}
\usepackage{enumerate}

\title{Renormalizing an initial state}
\author{Hael Collins, }
\author{R.~Holman, }
\author{Tereza Vardanyan}
\affiliation{Physics Department, Carnegie Mellon University, Pittsburgh PA 15213 USA}
\emailAdd{hcollins@andrew.cmu.edu}
\emailAdd{rh4a@andrew.cmu.edu}
\emailAdd{tvardany@andrew.cmu.edu}

\abstract{The intricate machinery of perturbative quantum field theory has largely been
devoted to the `dynamical' side of the theory:  simple states are evolved in
complicated ways.  This article begins to address this lopsided treatment.  Although
it is rarely possible to solve for the eigenstates of an interacting theory exactly,
a general state and its evolution can nonetheless be constructed perturbatively in
terms of the propagators and structures defined with respect to the free theory. 
The detailed form of the initial state in this picture is fixed by imposing suitable
`renormalization conditions' on the Green's functions.  This technique is
illustrated with an example drawn from inflation, where the presence of
nonrenormalizable operators and where an expansion that naturally couples early
times with short distances make the ability to start the theory at a finite initial
time especially desirable.}

\arxivnumber{}

\begin{document}
\maketitle

\setcounter{page}{2}

\section{Introduction}
For calculations in quantum field theory we usually start with the appropriate quadratic action, derive the propagator for this free theory and, based on it, construct Green's functions of the full theory perturbatively. The standard applications assume relatively simple states. In scattering problems, the ``in" and ``out" states are chosen to be the free theory single particle states in an infinite past and future. In inflationary calculations the ``in" state is the free Bunch-Davies state in an infinite past. This is what we do in practice. But in both cases we really mean to be in the eigenstate of the full theory. The reason why using the free eigenstates gives us the correct answer is because the states are being evolved over an infinite time. In this situation we can use mathematical tricks like an $i\epsilon$ prescription or an adiabatic switching on of the interaction to separate the full eigenstate from the free one. For example, the usual logic for calculating cosmological correlation functions in the vacuum state of an interacting theory is to start the evolution from an early enough time $t_0\to-\infty$. Then it is possible to argue that there are no contributions from the lower end of the time integrals: the fields oscillate rapidly, and after deforming the integration contour ($i\epsilon$ prescription) to project out the full vacuum these terms go to zero.

But let us say that we want to start our evolution from an arbitrary initial time; then we cannot use these procedures to pick out the vacuum state we want. Moreover, if we want to calculate correlation functions not in the full vacuum, but in some arbitrary state of an interacting theory, then even if we started from $-\infty$ we still will not be able to use the $i\epsilon$ prescription since it can only project out a state with the lowest energy, i.e.\ the vacuum state.

There are several reasons for wanting to start from a finite initial time. First of all, for a lot of states neither an $i\epsilon$ prescription, nor an adiabatic ``turning on'' of an interaction are useful, so there is no advantage in taking $t_0\to-\infty$. For instance, the state might not be an equilibrium state of the interacting theory. Starting in the infinite past and ``turning on'' the interactions, we will not naturally flow into such a state. Another example is a bound state in an interacting theory. This state will not exist in the infinite past once we have ``turned off'' the interactions. In this case something discrete happens: either particles are bound or they are not; there is no adiabatic transition between these two statements.

Secondly, we will be able to treat interesting excited states that might not necessarily have a reasonable extrapolation all the way back to $t_0\to-\infty$, but which are sensible enough (non-singular) at a finite time $t_0$. In this case it is really the state itself that is important, not the particular value of $t_0$ that we have chosen, as long as it remains finite, since we are not assuming that anything physical is happening at $t_0$.       

Thirdly, there is a danger that by going back to the infinite past we might enter a non-perturbative regime or a regime in which there might be some uncontrolled, poorly understood UV behavior as $t_0\to-\infty$. The trans-Planckian problem of inflation is an example of this case. Because of the expansion, going to the infinite past is equivalent to going to arbitrarily short distances. But we know that once we reach distances smaller than the Planck scale the contributions from higher order operators will become more and more important and we will end up having an infinite number of unsuppressed nonrenormalizable operators. Thus, we would like to be able to start our evolution from scales far enough from the Planck threshold.

And the last, but most obvious reason is that something is really happening at $t_0$, so it is a natural choice to use. 

In this paper we present a different approach for calculating the expectation values of the products of fields that can be applied in the case of a finite initial time. At this initial time let our fields be in some state, for example, the vacuum state, a thermal state, etc. We can construct such a state through a set of boundary operators on the initial time hypersurface \cite{Agarwal:2012mq}. These operators are implicitly defined with respect to the free theory vacuum. However, what we really want is to calculate correlation functions of an interacting theory in the corresponding interacting theory state, e.g.\  the interacting vacuum, an interacting thermal state, etc. Therefore we need to renormalize the structures of the initial state perturbatively, order by order in the parameters of the interacting theory, in such a way that this initial state satisfies certain conditions. This is somewhat similar to how operators are renormalized in the dynamical part of a Lagrangian in ordinary quantum field theory. We know how certain $n$-point functions behave in the free theory case; for example, we know that the one-point function is zero and the pole of the propagator has a residue of 1, and we would like to have the same behavior for these functions in the full theory. As a consequence of imposing this behavior we have to rescale fields and introduce counterterms.   

This renormalization is required even in the simplest case---an interacting theory in its vacuum state at a particular time $t_0\neq -\infty$. We find that the corrections to the $n$-point functions have an explicit dependence on the initial time. When taking $t_0\to-\infty$ we see that these functions do not match to the ones that we get when we start evolving from the asymptotic vacuum: they contain additional divergent and oscillatory terms. This means that at $t_0$ we were in the wrong state, not in the state we intended to be, i.e.\ not in the interacting vacuum state. To fix this we add operators and structures to the initial state action---these are the ``counterterms" of this picture, and they are defined order by order.    
    
In the next section we will show how to specify order by order in perturbation theory the initial state using the eigenstates of the free part of the theory. Section 3 mentions a few details of simple single-field, slow-roll inflationary models that will be used in our calculations. Sections 4 and 5 are the sample calculations of the vacuum state three- and two-point functions of inflation using this method and the fact that we know what we should get for $t_0\to-\infty$ from the conventional calculations.

\section{Changing bases}
Let the operator ${\cal O}$ be a product of fields. In the Schr\"odinger picture its expectation value at a time $t$ is given by
\begin{equation}
\langle{\cal O}\rangle(t)\equiv\langle\Omega(t_0)|U^\dagger(t,t_0){\cal O}U(t,t_0)|\Omega(t_0)\rangle\;,
\end{equation}
where $|\Omega(t_0)\rangle$ is the state of the system at the initial time $t_0$. 
The time-evolution operator $U(t,t_0)$ satisfies the Schr\"odinger equation
\begin{equation}
i\frac{d}{dt}U(t,t_0) = H(t)\, U(t,t_0)\;
\end{equation}
with $U(t_0,t_0)={\mathbb I}$ as the initial condition. Here $H(t)$ is the full Hamiltonian of the system.

Suppose that at $t_0$ the system was in its vacuum state, i.e. $|\Omega(t_0)\rangle$ is such that $E_0\equiv\langle\Omega(t_0)|H(t_0)|\Omega(t_0)\rangle$ is the lowest energy assumed by any state at $t_0$. In most cases we are not able to find the explicit form of the full vacuum $|\Omega(t_0)\rangle$, but usually we can solve for the eigenstates of a part of the Hamiltonian, which we call $H_0$ and which corresponds to the free part of the theory,
\begin{equation}
H(t) = H_0(t) + H'(t) .
\end{equation}
Let us suppose that we have solved the eigenvalue problem for $H_0(t_0)$ at the initial time.  The set of eigenstates of $H_0(t_0)$ can be used as a basis of our Hilbert space. We label them as 
$$
\bigl\{ |0(t_0)\rangle, |n(t_0)\rangle\bigr\} .
$$
The state $|0(t_0)\rangle$ denotes the vacuum state of the free theory at $t_0$, and $|n(t_0)\rangle$ collectively represents all of the other eigenstates of $H_0$. We assume that this is a complete set in the sense that we can expand the identity operator in terms of it
$$
{\mathbb I} = |0(t_0)\rangle\langle0(t_0)| 
+ \sum_n |n(t_0)\rangle\langle n(t_0)| . 
$$
We can use this completeness relation to convert a state in the eigenbasis of the full theory into its expression in the free theory's eigenbasis. The density matrix of the initial state $\rho_0=\rho(t_0) = |\Omega(t_0)\rangle\langle\Omega(t_0)|$ can be written as
\begin{eqnarray} 
\rho_0 
= {\mathbb I}\, |\Omega(t_0)\rangle\langle\Omega(t_0)| \, {\mathbb I} 
&=& 
|0(t_0)\rangle\, \langle0(t_0)|\Omega(t_0)\rangle\langle\Omega(t_0)|0(t_0)\rangle\, \langle0(t_0)|
\nonumber \\
&& 
+\sum_n |n(t_0)\rangle\, \langle n(t_0)|\Omega(t_0)\rangle\langle\Omega(t_0)|0(t_0)\rangle\, \langle0(t_0)|
\nonumber \\
&& 
+\sum_n |0(t_0)\rangle\, \langle 0(t_0)|\Omega(t_0)\rangle\langle\Omega(t_0)|n(t_0)\rangle\, \langle n(t_0)|
\nonumber \\
&& 
+\sum_{n,n'} |n(t_0)\rangle\, \langle n(t_0)|\Omega(t_0)\rangle\langle\Omega(t_0)|n'(t_0)\rangle\, \langle n'(t_0)|.
\nonumber 
\end{eqnarray}
In general, $\langle n(t_0)|\Omega(t_0)\rangle\not=0$, which means that from the perspective of the free theory, the true vacuum state looks as though it contains multiparticle excitations.  But that is only because we are using the ``wrong'' basis; in the basis of the eigenstates of the full theory, $|\Omega(t_0)\rangle$ does not contain any excitations.  It is the lowest energy state.

We have been speaking as though we knew $|\Omega(t_0)\rangle$, $U(t,t_0)$, etc.  But if we did, there would be no need ever to resort to the eigenstates of the free theory.  So how do we proceed, not knowing $\rho_0$?  Let us make a few observations:  
\vskip-18truept $\quad$

\begin{enumerate}[(1)]
\addtolength{\itemsep}{-6pt}
\item If we really knew $\rho_0$ in the free eigenbasis, then we could calculate the expectation values of any operator (in principle) in the full vacuum state.  Therefore, we should try to determine $\rho_0$ in this basis somehow.  
\item $\rho_0$---even though it is a pure state in the full eigenbasis---is a {\it mixed state\/} in the free theory's eigenbasis; that is,
$$
\rho_{nn'}= \langle n(t_0)|\Omega(t_0)\rangle\langle\Omega(t_0)|n'(t_0)\rangle 
$$
does not need to be diagonal.
\end{enumerate}
\vskip-6truept

So the problem that we wish to solve is to evaluate an operator in a basis that we {\it do\/} understand with an initial state that we {\it do not\/} know.  When $H'(t)$ is ``small'' in some sense, we can evaluate the expectation value perturbatively.  In fact our approach will be perturbative in a double sense.  First, by dividing $H=H_0+H'$, we can similarly divide the time-evolution operator, $U(t,t_0) = U_0(t,t_0)U_I(t,t_0)$.  Thus, we can write the expectation value of $\cal{O}$ in the interaction picture as
\begin{eqnarray} 
\langle{\cal O}(t)\rangle &=& 
{\rm tr}\, \bigl[ U_I^\dagger(t,t_0) U_0^\dagger(t,t_0) {\cal O} 
U_0(t,t_0)U_I(t,t_0) \rho_0 \bigr]
\nonumber \\
&\!\!\!\!\!\!=\!\!\!\!\!\!& 
{\rm tr}\, \bigl[ U_I^\dagger(t,t_0) {\cal O}_I(t)U_I(t,t_0) \rho_0 \bigr]
\nonumber 
\end{eqnarray}
where ${\cal O}_I(t) = U_0^\dagger(t,t_0) {\cal O}U_0(t,t_0)$ is the operator ${\cal{O}}$ in the interaction picture and $U_0(t,t_0) = T e^{-i\int_{t_0}^t dt'\, H_0(t')}$. The idea is that if $H'$---or the corresponding interaction Hamiltonian in the interaction picture $H_I= U_0^\dagger(t,t_0) H' U_0(t,t_0)$---is small, we can treat the interactions pertubatively by expanding 
$$
U_I(t,t_0) = T e^{-i\int_{t_0}^t dt'\, H_I(t')} 
$$
in powers of $H_I$.

The second perturbative expansion is based on the idea that if $H$ is close to $H_0$, $|0(t_0)\rangle$ ought also to be ``close to'' $|\Omega(t_0)\rangle$ in the sense that the overlap with the multi-particle states is small.  If we can establish a few suitable criteria, we can determine $\rho_0$ in the free theory eigenbasis {\it perturbatively\/}.  For example, 
\vskip-18truept $\quad$

\begin{enumerate}[(1)]
\addtolength{\itemsep}{-6pt}
\item $\rho_0$ should have the same symemtries as the full vacuum.  
\item If we believe that the state should match with what we should have obtained by extending back to the $t_0\to-\infty$, then that requires certain structures in $\rho_0$.
\end{enumerate}
\vskip-6truept

The only variables around are the fields $\zeta(t,\vec{x})$; therefore, we should have that $\rho_0=\rho(\zeta(t_0,\vec{x});t_0)$. It is convenient to write the initial density matrix in the following general form
\begin{equation*}
\rho_0=\frac{1}{Z}e^{iS_0}\;,
\end{equation*}
where $Z$ is such that ${\rm tr}(\rho_0)=1$. This idea was introduced in \cite{Agarwal:2012mq}. Since a particular configuration of the fields at the initial time $t_0$ is then weighted by a $e^{iS_0}$ factor, we can think of $S_0$ as a boundary action on the initial time hypersurface \cite{Collins:2013kqa}. Hence, the problem of determining the initial density matrix is reduced to the problem of constructing an appropriate initial action.

\section{Single field inflation}
Let us use the method we described in the previous section to calculate several cosmological correlation functions. We will work with a simple single-field, slow-roll inflationary model whose action is given by
\begin{equation*}
S=\int d^4x\;\sqrt{-g}\left\{\frac{1}{2}M_{pl}^2R+\frac{1}{2}g^{\mu\nu}\partial_{\mu}\phi\partial_{\nu}\phi-V(\phi)\right\}\;.
\end{equation*}
The metric for the spatially invariant background can be written as
\begin{equation*}
{ds}^2={dt}^2-e^{2\rho(t)}\delta_{ij}dx^idx^j\;.
\end{equation*}
To analyze the fluctuations about this background it is convenient to write the metric in the following form
\begin{equation*}
{ds}^2=\left[N^2-h_{ij}N^iN^j\right]{dt}^2-2h_{ij}N^idtdx^j-h_{ij}dx^idx^j\;.
\end{equation*}
Choosing the coordinates in which there are no fluctuations in the inflaton field $\phi(t,\vec{x})=\phi(t)$ and where the spatial part of the metric is proportional to $\delta_{ij}$ and neglecting the tensor fluctuations we can write that
\begin{equation*}
h_{ij}=e^{2\rho(t)+2\zeta(t,\vec{x})}\delta_{ij}\;.
\end{equation*}
In these coordinates the only scalar fluctuation left is $\zeta(t,\vec{x})$. The quadratic part of its action is 
\begin{equation}
S^{(2)}=\frac{1}{2}\int dt\;\frac{{\dot{\phi}}^2}{{\dot{\rho}}^2}\int d^3\vec{x}\;e^{3\rho(t)}\left\{{\dot{\zeta}}^2-e^{-2\rho(t)}\partial_k\zeta\partial^k\zeta\right\}\label{s2}\;.
\end{equation}
The fields $N$ and $N^i$ are both nondynamical Lagrange multipliers, satisfying constraint equations
\begin{equation*}
N=1+\frac{\dot\zeta}{\dot\rho}\;,
\end{equation*}
\begin{equation*}
N^i=\delta_{ij}\partial^j\left\{-\frac{e^{-2\rho}}{\dot\rho}\zeta+\frac{1}{2}\frac{{\dot\phi}^2}{{\dot\rho}^2}\partial^{-2}\dot\zeta\right\}\;.
\end{equation*}

Expanding the inflationary action to third order in $\zeta(t,\vec{x})$ and going through lots of lengthy manipulations, in particular, doing many integrations by parts, the cubic action can be put into the following form \cite{Maldacena:2002vr,Collins:2011mz} 
\begin{eqnarray}
S^{(3)}&=&M_{pl}^2\int d^4x\;\bigg\{\epsilon(3\epsilon+2\delta)e^{\rho}\zeta\partial_k\zeta\partial^k\zeta-\epsilon(\epsilon+2\delta)e^{3\rho} {\dot{\zeta}}^2\zeta-2\epsilon^2e^{3\rho}\dot{\zeta}\partial_k\zeta\partial^k\zeta\partial^{-2}\dot{\zeta}\nonumber\\&&\qquad\qquad\quad-\frac{1}{2}e^{3\rho}\epsilon^3[{\dot\zeta}^2\zeta-\zeta\partial_k\partial_l(\partial^{-2}\dot\zeta)\partial^k\partial^l(\partial^{-2}\dot\zeta)]\nonumber\\&&\qquad\qquad\quad+\Big\{\frac{d}{dt}[\epsilon e^{3\rho}\dot\zeta]-\epsilon e^{\rho}\partial_k\partial^k\zeta\Big\}\Big\{\frac{2}{\dot\rho}\dot\zeta\zeta-\frac{1}{2}\frac{e^{-2\rho}}{{\dot\rho}^2}[\partial_k\zeta\partial^k\zeta-\partial^{-2}\partial_k\partial_l(\partial^k\zeta\partial^l\zeta)]\nonumber\\&&\qquad\qquad\qquad\qquad\qquad\qquad\qquad\qquad+\frac{1}{\dot\rho}\epsilon[\partial_k\partial^k(\partial^{-2}\dot\zeta)-\partial^{-2}\partial_k\partial_l(\partial^k\zeta\partial^l(\partial^{-2\dot\zeta}))]\Big\}\bigg\}\;,\nonumber\\\label{s3}
\end{eqnarray}
where $\epsilon$ and $\delta$ are small in the slow-roll limit
\begin{equation*}
\epsilon=\frac{1}{2}\frac{1}{M_{pl}^2}\frac{{\dot{\phi}}^2}{{\dot{\rho}}^2}\ll1\;,
\end{equation*}
\begin{equation*}
\delta=\frac{1}{H}\frac{\ddot{\phi}}{\dot{\phi}}\ll1\;.
\end{equation*}
Only the first three operators in \eqref{s3} have contributions that don't vanish in the late-time limit.

\section{The three-point function}
For simplicity, we will analyze the correlation functions using an abbreviated version of the standard single-field inflationary theory. We use the quadratic action given in \eqref{s2}, but from among the operators in the cubic action we will be only looking at one,
\begin{equation*}
S^{(3)}=\int d^4x\;M_{pl}^2\left\{\epsilon(3\epsilon+2\delta)e^{\rho(t)}\zeta\partial_k\zeta\partial^k\zeta\right\}\;.
\end{equation*}
There are two reasons for doing so. First of all, for what we are trying to illustrate here, adding more cubic terms will not be any more illuminating and will only lengthen and complicate the calculation. Secondly, the standard ``late-time'', leading slow-roll set of operators is not even sufficient if we really wish to renormalize the single-field inflationary model. The renormalization must be done at an arbitrary time and not just in the late-time limit. All of the operators in \eqref{s3} must be included then. 

To calculate the three-point function here, and the two-point function in the next section, we work in the interaction picture and use the ``in-in'' formalism \cite{Schwinger:1960qe}. In this formalism the three-point function can be written as   
\begin{eqnarray}
&&\!\!\!\!\!\!\!\!\!\!\!\!\!\!\!\!\!\!\!\!\!\!
\langle \Omega(t)|\zeta(t,\vec{x})\zeta(t,\vec{y})\zeta(t,\vec{z})|\Omega(t) \rangle\nonumber\\ &=&\langle \Omega(t_0)|U_I^\dagger(t,t_0)\zeta(t,\vec{x})\zeta(t,\vec{y})\zeta(t,\vec{z})U_I(t,t_0)|\Omega(t_0) \rangle\nonumber\\ & = & \langle\Omega(t_0)|T(\zeta^+(t,\vec{x})\zeta^+(t,\vec{y})\zeta^+(t,\vec{z})e^{-i\int_{t_0}^t dt'\;[H_I^+(t')-H_I^-(t')]})|\Omega(t_0) \rangle\nonumber\\ & = & -i\int_{t_0}^t dt'\;\langle \Omega(t_0)|T(\zeta^+(t,\vec{x})\zeta^+(t,\vec{y})\zeta^+(t,\vec{z})[H_I^+(t')-H_I^-(t')])|\Omega(t_0) \rangle+\cdots\;, \label{zero}
\end{eqnarray}
where
\begin{equation}
H_I(t)=-M_{pl}^2\epsilon(3\epsilon+2\delta) e^{\rho(t)}\int d^3\vec{x}\;\zeta\partial_k\zeta\partial^k\zeta\label{HI}\;
\end{equation}
and $H_I^{\pm}(t)\equiv H_I^+[\zeta^{\pm}(t,\vec{x})]$. The fields $\zeta^+(t,\vec{x})$ and $\zeta^-(t,\vec{x})$ are associated with $U_I(t,t_0)$ and $U_I^\dagger(t,t_0)$ respectively. The time-ordering operation is extended in the following sense: two ``+" fields are ordered in the usual way,
\begin{equation*}
T(\zeta^+(t,\vec{x})\zeta^+(t',\vec{y}))=\Theta(t-t')\zeta^+(t,\vec{x})\zeta^+(t',\vec{y})+\Theta(t'-t)\zeta^+(t',\vec{y})\zeta^+(t,\vec{x})\;,
\end{equation*}
``--" fields always occur after ``+" fields,
\begin{eqnarray*}
T(\zeta^+(t,\vec{x})\zeta^-(t',\vec{y}))&=&\zeta^-(t',\vec{y})\zeta^+(t,\vec{x})\;,\\
T(\zeta^-(t,\vec{x})\zeta^+(t',\vec{y}))&=&\zeta^-(t,\vec{x})\zeta^+(t',\vec{y})\;,
\end{eqnarray*}
and two ``--" fields are ordered in the opposite of the usual sense,
\begin{equation*}
T(\zeta^-(t,\vec{x})\zeta^-(t',\vec{y}))=\Theta(t'-t)\zeta^-(t,\vec{x})\zeta^-(t',\vec{y})+\Theta(t-t')\zeta^-(t',\vec{y})\zeta^-(t,\vec{x})\;.
\end{equation*}
Correspondingly, there are four types of propagators
\begin{eqnarray*}
\langle \Omega(t_0)|T(\zeta^+(t,\vec{x})\zeta^+(t',\vec{y}))|\Omega(t_0)\rangle&=&G^{++}(t,\vec x;t',\vec y)=\Theta(t-t')G^{>}(t,\vec x;t',\vec y)+\Theta(t'-t)G^{<}(t,\vec x;t',\vec y)\;,\\
\langle \Omega(t_0)|T(\zeta^+(t,\vec{x})\zeta^-(t',\vec{y}))|\Omega(t_0)\rangle&=&G^{+-}(t,\vec x;t',\vec y)=G^{<}(t,\vec x;t',\vec y)\;,\\
\langle \Omega(t_0)|T(\zeta^-(t,\vec{x})\zeta^+(t',\vec{y}))|\Omega(t_0)\rangle&=&G^{-+}(t,\vec x;t',\vec y)=G^{>}(t,\vec x;t',\vec y)\;,\\
\langle \Omega(t_0)|T(\zeta^-(t,\vec{x})\zeta^-(t',\vec{y}))|\Omega(t_0)\rangle&=&G^{--}(t,\vec x;t',\vec y)=\Theta(t'-t)G^{>}(t,\vec x;t',\vec y)+\Theta(t-t')G^{<}(t,\vec x;t',\vec y)\;.
\end{eqnarray*}
Here $G^{>}(t,\vec x;t',\vec y)$ and $G^{<}(t,\vec x;t',\vec y)$ are Wightman functions
\begin{eqnarray*}
G^{>}(t,\vec x;t',\vec y)&=&\langle \Omega(t_0)|\zeta(t,\vec{x})\zeta(t',\vec{y}))|\Omega(t_0)\rangle=\int\frac{d^3\vec{k}}{(2\pi)^3}\;e^{i{\vec{k}(\vec{x}-\vec{y})}}G_k^{>}(t,t')\;,\\
G^{<}(t,\vec x;t',\vec y)&=&\langle \Omega(t_0)|\zeta(t',\vec{y})\zeta(t,\vec{x}))|\Omega(t_0)\rangle=\int\frac{d^3\vec{k}}{(2\pi)^3}\;e^{i{\vec{k}(\vec{x}-\vec{y})}}G_k^{<}(t,t')\;.
\end{eqnarray*}
Using these rules to perform the contractions in \eqref{zero}, we find that the leading contribution to the three-point function is
\begin{eqnarray}
&&\!\!\!\!\!\!\!\!\!\!\!\!\!\!\!\!
\langle\zeta_{\vec k_1}(t)\zeta_{\vec k_2}(t)\zeta_{\vec k_3}(t)\rangle 
\nonumber \\
& = & -2iM^2\epsilon(3\epsilon+2\delta)[\vec k_1\cdot\vec k_2+\vec k_1\cdot\vec k_3+\vec k_2\cdot\vec k_3]\nonumber\\&&\quad\times\int_{t_0}^t dt\;e^{\rho(t')}\;\left\{G_{k_1}^>(t,t')G_{k_2}^>(t,t')G_{k_3}^>(t,t')- G_{k_1}^<(t,t')G_{k_2}^<(t,t')G_{k_3}^<(t,t')\right\}\;.\quad
\label{one}
\end{eqnarray}
Since $\vec k_1+\vec k_2+\vec k_3=0$, we can rewrite the coefficients in a form that only depends on the magnitudes of the momenta,
\begin{equation*}
\vec k_1\cdot\vec k_2+\vec k_1\cdot\vec k_3+\vec k_2\cdot\vec k_3=-\frac{1}{2}[k_1^2+k_2^2+k_3^2]\;.
\end{equation*}
To evaluate the time integral, let us switch from $t$ to the conformal time $\eta$. Since we are working at leading order in the slow-roll parameters, we can write the scale factor and the integration measure in the de Sitter limit, 
\begin{equation*}
\int_{t_0}^t dt'\;e^{\rho(t')}\cdots=\int_{\eta_0}^{\eta} d\eta'\;\frac{dt'}{d\eta'}\;e^{\rho(t')}\cdots=\int_{\eta_0}^{\eta} d\eta'\;e^{2\rho(t')}\cdots=\int_{\eta_0}^{\eta} d\eta'\;\frac{1}{H^2\eta'^2}\cdots\;.
\end{equation*}
In the standard case, where $t_0\to-\infty$, on the right-hand side of \eqref{zero} one replaces $|\Omega(t_0)\rangle$ with the vacuum state of the free theory $|0\rangle\equiv|0(t_0)\rangle$, which in practice means using the Wightman functions of the free theory to evaluate \eqref{one}. Then $t_0$ is set to $-\infty(1\pm i\epsilon)$ to project out the vacuum state of the interacting theory $|\Omega(t_0)$ from the vacuum state of the free theory $|0\rangle$. The Wightman functions of the free theory associated with the Bunch-Davies vacuum are
\begin{eqnarray}
G_k^>(t,t')&=&\frac{1}{4\epsilon}\frac{H^2}{M_{pl}^2}\frac{1}{k^3}(1+ik\eta)(1-ik\eta')e^{-ik(\eta-\eta')}\nonumber\\
G_k^<(t,t')&=&\frac{1}{4\epsilon}\frac{H^2}{M_{pl}^2}\frac{1}{k^3}(1-ik\eta)(1+ik\eta')e^{-ik(\eta-\eta')}\;. \label{W}
\end{eqnarray}
Substituting \eqref{W} into \eqref{one} and using the $i\epsilon$ prescription, which gets rid of the terms coming from the lower limit of the integral, we find that the three-point function is equal to
\begin{eqnarray}
\langle\zeta_{\vec{k_1}}(t)\zeta_{\vec{k_2}}(t)\zeta_{\vec{k_3}}(t)\rangle &=&\frac{(3\epsilon+2\delta)}{32\epsilon^2}\frac{H^4}{M^4}\frac{(k_1^2+k_2^2+k_3^2)}{k_1^3k_2^3k_3^3}
\label{two} \\
&&\times
\biggl\{K-\frac{k_1k_2+k_1k_3+k_2k_3}{K}-\frac{k_1k_2k_3}{K^2}
\nonumber \\
&&\quad
+\biggl( \frac{(k_1k_2+k_1k_3+k_2k_3)^2}{K}+\frac{(k_1k_2+k_1k_3+k_2k_3)k_1k_2k_3}{K^2} \biggr)\eta^2+\frac{k_1^2k_2^2k_3^2}{K}\eta^4\biggr\}\;,
\nonumber 
\end{eqnarray}
where
\begin{equation*}
K=k_1+k_2+k_3\;.
\end{equation*}

But what should we do when $t_0$ is finite? Let us once again try to use \eqref{W} as our Wightman functions. In this instance, one part of the three-point function is the same as in \eqref{two}, but there is also a piece from the lower limit of the integral in \eqref{one}, which is equal to
\begin{multline}
\frac{(3\epsilon+2\delta)}{32\epsilon^2}\frac{H^4}{M^4}\frac{(k_1^2+k_2^2+k_3^2)}{k_1^3k_2^3k_3^3}\Big\{\frac{k_1k_2k_3}{K^2}[A\cos K(\eta_0-\eta)+B\sin K(\eta_0-\eta)]\\
\quad\quad\quad\quad\quad\quad\quad\quad\quad\quad\quad\quad\quad+\frac{k_1k_2+k_1k_3+k_2k_3}{K}[A\cos K(\eta_0-\eta)+B\sin K(\eta_0-\eta)] \\
\quad\quad\quad\quad\quad\quad\quad\quad\quad+\frac{k_1k_2k_3}{K}\eta_0[B\cos K(\eta_0-\eta)+A\sin K(\eta_0-\eta)] \\+\frac{1}{\eta_0}[B\cos K(\eta_0-\eta)+A\sin K(\eta_0-\eta)]\Big\}\label{three}\;,
\end{multline}
where 
\begin{eqnarray*}
A &=& 1-(k_1k_2+k_1k_3+k_2k_3)\eta^2 \\
B &=& K\eta-k_1k_2k_3\eta^3\;. 
\end{eqnarray*}
There are terms in \eqref{three} that either diverge or remain finite as $\eta_0\to-\infty$. The reason for the appearance of these terms is the fact that the free Bunch-Davies state is not the vacuum of the interacting theory. Since we are starting our evolution from a finite $t_0$, we can't simply use the $i\epsilon$ prescription to project out the vacuum state of the full theory. However, if we want to match smoothly with the interacting vacuum in the $\eta_0\to-\infty$ limit, another recourse is open to us: to put a cubic term in the initial action. From what we have said earlier, this is equivalent to modifying the initial state, described in terms of the basis of the free theory at $t_0$, so that it corresponds more closely to the state that we really intended it to be. To do so we use a boundary operator whose structure mirrors the structure of $S^{(3)}$, 
\begin{eqnarray}
S_0^{(3)} &=& M^2\epsilon(3\epsilon+2\delta) e^{2\rho(t_0)}\int d^3\vec{x}\;d^3\vec{y}\;d^3\vec{z}\bigl\{C(\vec{x},\vec{y},\vec{z})\zeta^+(t_0,\vec{x})\partial_k\zeta^+(t_0,\vec{y})\partial^k\zeta^+(t_0,\vec{z})\nonumber\\&&\qquad\qquad\qquad\qquad\qquad\qquad\qquad\qquad-C^*(\vec{x},\vec{y},\vec{z})[\zeta^+\to\zeta^-]\bigr\}\;.\nonumber\\&&\label{s03}
\end{eqnarray}
For this surface action to cancel the unwanted terms, we need
\begin{equation*}
C(\vec{k_1},\vec{k_2},\vec{k_3})=\frac{1}{K}\Big(\frac{1}{K\eta_0}-i\Big)\;.
\end{equation*}
By using $S^{(3)}+S_0^{(3)}$ as our cubic action to calculate the three-point function for a general $t_0$ we will recover \eqref{two} when taking $t_0\to-\infty$. Notice, that for $t_0\neq-\infty$, the three-point function will not be equal to \eqref{two}. It will have some additional pieces that depend on $t_0$, but they all vanish when $t_0\to-\infty$.

\section{A one-loop correction to the two-point function}
If we try to evaluate the two-point function beyond leading order with a finite time, we encounter the same problem as occurred with the three-point function: the lower ends of the integrals associated with the time-evolution of the states will produce pieces that are finite but oscillatory or that are divergent as we take $t_0\to-\infty$. But here we should be more careful when removing these terms. The reason is that in this case there are other divergences coming from the dynamical part itself: the divergences of the three-momentum integrals in the loop. To take care of them we must supply the usual counterterms in the Lagrangian. These in turn will affect the initial time dependence of the two-point function. Only once we have summed both loop and the counterterm graphs, and  isolated the finite oscillatory and divergent parts as $t_0\to -\infty$ will we be able to determine the appropriate way to modify the state to cancel these effects.

\subsection{Renormalizing the standard vacuum state}
Using the ``in-in" formalism we can write the two-point function as
\begin{eqnarray}
\!\!\!\!\!\!\!\!\!\!\langle\Omega(t)|\zeta(t,\vec{x})\zeta(t,\vec{y})|\Omega(t) \rangle=\langle\Omega(t_0)|T(\zeta^+(t,\vec{x})\zeta^+(t,\vec{y})e^{-i\int_{t_0}^t dt'\;[H_I^+(t')-H_I^-(t')]})|\Omega(t_0) \rangle\; .\label{2point}
\end{eqnarray}
 For the one-loop contribution we have
\begin{eqnarray*}
\lefteqn{-\frac{1}{2}\int_{t_0}^t dt' \int_{t_0}^t dt''\;\langle\Omega(t_0)|T(\zeta^+(t,\vec{x})\zeta^+(t,\vec{y})[H_I^+(t')-H_I^-(t')])[H_I^+(t'')-H_I^-(t'')])|\Omega(t_0) \rangle}\nonumber\\&=&-M_{pl}^4\epsilon^2(3\epsilon+2\delta)^2\int\frac{d^3\vec{p}}{(2\pi)^3}\;e^{i{\vec{p}(\vec{x}-\vec{y})}}\int_{t_0}^t dt'\;e^{\rho(t')}\int_{t_0}^{t'} dt''\;e^{\rho(t'')}\{G_p^>(t,t')-G_p^<(t,t')\}\nonumber\\&&\times\int\frac{d^3\vec{q}}{(2\pi)^3}\;(p^2+q^2+k^2)^2\{  G_p^>(t,t'')G_q^>(t',t'')G_k^>(t',t'')- G_p^<(t,t'')G_q^<(t',t'')G_k^<(t',t'')\}\;, 
\end{eqnarray*}
where
\begin{eqnarray*}
k=|\vec{p}-\vec{q}|\;.
\end{eqnarray*}
Again, for the case where $t_0=-\infty$ we use the free Bunch-Davies Wightman functions and the $i\epsilon$ prescription for the lower ends of both integrals. Then the zeroth order contribution is just the usual Bunch-Davies propagator and the one-loop contribution is equal to
\begin{equation}
\langle\zeta_{\vec{p}}(t)\zeta_{-\vec{p}}(t)\rangle_{\rm loop}=\frac{(3\epsilon+2\delta)^2}{256\epsilon^2}\frac{H^4}{M^4}\frac{1}{p^3}\{I_0+p^2\eta^2 I_2+p^4\eta^4 I_4\}\label{BD1}\;,
\end{equation}
where
\begin{eqnarray*}
I_0 & = &\frac{1}{2p^4}\int\frac{d^3\vec{q}}{(2\pi)^3}\;\frac{(p^2+q^2+k^2)^2}{q^3k^3(p+q+k)^2}\nonumber\\&&\times\{4kp^2(2p^2+2pq+3q^2)+2p^2(p+q)(2p^2+2pq+3q^2)\nonumber\\&& \quad+k^3(6p^2+5q^2)+k^2(10p^3+12p^2q+8pq^2+5q^3)\}\;,\\\\
I_2 & = & \frac{1}{2p^4}\int\frac{d^3\vec{q}}{(2\pi)^3}\;\frac{(p^2+q^2+k^2)^2}{q^3k^3(p+q+k)^2}
\nonumber\\
&&\times\{4kp^2q^2+2p^2q^2(p+q)+k^3(2p^2+5q^2)+k^2(2p^3+4p^2q+8pq^2+5q^3)\}\;,\\\\
I_4 & = &\frac{1}{p^4}\int\frac{d^3\vec{q}}{(2\pi)^3}\;\frac{(p^2+q^2+k^2)^2}{q^3k^3(p+q+k)^2}\{k^2q^2(k+2p+q)\}\;.
\end{eqnarray*}
By doing power counting we can see that these integrals have divergences. In order to remove them, we introduce the necessary counterterms,
\begin{equation*}
{\cal L}_{ct}=c_1 M_{pl}^2e^{3\rho(t)}\dot{\zeta^2}-c_2 M_{pl}^2e^{\rho(t)}\partial_k\zeta\partial^k\zeta-c_3e^{-\rho(t)}\partial_l\partial_k\zeta\partial^l\partial^k\zeta\;.
\end{equation*}
The first two counterterms, which renormalize the operators in the quadratic action \eqref{s2}, are not enough to remove all divergences. We need the last four-derivative operator to cancel divergences proportional to $p^4\eta^4$. The $e^{-\rho(t)}$ prefactor is the one appropriate for the geometry: each pair of spatial indices is contracted with an $h^{ij}$, each of which brings an $e^{-2\rho(t)}$, and there is an overall factor of $\sqrt{-g}$ from the coordinate-invariant measure, which brings $e^{3\rho(t)}$. 
The corresponding contributions from these counterterms to the two-point function are
\begin{equation*}
-\frac{1}{8\epsilon^2}\frac{H^2}{M^2}\frac{c_1}{p^3}(p^2\eta^2-1)\;,
\end{equation*}
\begin{equation*}
-\frac{1}{8\epsilon^2}\frac{H^2}{M^2}\frac{c_2}{p^3}(p^2\eta^2+3)\;,
\end{equation*}
\begin{equation*}
-\frac{1}{8\epsilon^2}\frac{H^4}{M^4} \frac{c_3}{2p^3}(2p^4\eta^4+5p^2\eta^2+5)\;.
\end{equation*}
To cancel the divergences due to the loop we should choose the coefficients of the counterterms to be
\begin{eqnarray*}
c_3 &=&\frac{(3\epsilon+2\delta)^2}{32}\Big[\hbox{infinite part of $I_4$}\Big]\;,\\
c_2 &=&\frac{H^2}{M^2}\frac{(3\epsilon+2\delta)^2}{128}\Big[\hbox{infinite part of $(I_0+I_2-5I_4)$}\Big]\;,\\ 
c_3 &=&\frac{H^2}{M^2}\frac{(3\epsilon+2\delta)^2}{128}\Big[\hbox{infinite part of $(3I_2-5I_4-I_0)$}\Big]\;.
\end{eqnarray*}
Hence, for the renormalized loop we have
\begin{equation}
\langle\zeta_{\vec{p}}(t)\zeta_{-\vec{p}}(t)\rangle_{\rm loop}^{\rm ren}=\frac{(3\epsilon+2\delta)^2}{256\epsilon^2}\frac{H^4}{M^4}\frac{1}{p^3}\{I_0^f+p^2\eta^2 I_2^f+p^4\eta^4 I_4^f\}\label{BD2}\;,
\end{equation}
where the $I^f$-s are the finite parts of the corresponding integrals.

\subsection{Renormalizing the vacuum state with an initial time}
To evaluate the correction to the two-point function in the case of a finite $t_0$ we first replace \eqref{2point} with its renormalized form,
\begin{eqnarray}
\!\!\!\!\!\!\!\!\!\!\langle\Omega(t)|\zeta(t,\vec{x})\zeta(t,\vec{y})|\Omega(t) \rangle=\langle\Omega(t_0)|T(\zeta^+(t,\vec{x})\zeta^+(t,\vec{y})e^{-i\int_{t_0}^t dt'\;[\bar H_I^+(t')-\bar H_I^-(t')]})|\Omega(t_0) \rangle\;,\label{2ren}
\end{eqnarray}
where 
\begin{equation*}
\bar{H}_I(t)=H_I(t)+H_{ct}(t)
\end{equation*}
and 
\begin{equation*}
H_{ct}=-{\cal L}_{ct}\;.
\end{equation*}
To be able to use the free theory Wightman functions we must switch from $|\Omega(t_0)\rangle$ to $|0\rangle$. When making this transition we need to take into account that from the perspective of the free theory the evolution is governed not just by the Hamiltonian $\bar H_I$, but also by the initial state cubic action \eqref{s03} that we already included to correct the three-point function. This means that we can replace the right-hand side of \eqref{2ren} with
\begin{equation}
 \langle 0|T(\zeta^+(t,\vec{x})\zeta^+(t,\vec{y})e^{-i\int_{t_0}^t dt'\;[\bar H_I^+(t')-\bar H_I^-(t')]+iS_0^{(3)}})|0 \rangle\label{fin0}\;.
\end{equation}
 Since $S_0^{(3)}$ is of the same order in the slow-role parameters as $H_I$ we need to take its contribution into account. Thus, the one-loop correction to the two-point function will be
\begin{equation}
-\frac{1}{2}\int_{t_0}^t dt' \int_{t_0}^t dt''\;\langle 0|T(\zeta^+(t,\vec{x})\zeta^+(t,\vec{y})[\bar{H}_I^+(t')-\bar{H}_I^-(t')+H_0^{(3)}(t')])[H_I^+(t'')-H_I^-(t'')+H_0^{(3)}(t'')])|0\rangle\label{fin1}\;, 
\end{equation}
where
\begin{equation*}
H_0^{(3)}(t)=-\frac{1}{2}\delta(t-t_0)S_0^{(3)}\;.
\end{equation*}
The part of \eqref{fin1} that is independent of the initial time $\eta_0$ will be the same as \eqref{BD2}. The part that depends on $\eta_0$ will have terms that vanish, stay finite (and oscillate) or diverge (linearly and quadratically in $\eta_0$) as $\eta_0\to-\infty$. But when $\eta_0\to-\infty$ we want \eqref{fin1} to match with \eqref{BD2}; hence, we need to eliminate the last two types of terms. It can be done order by order in $\eta_0$. Here we will present the elimination of the quadratically divergent terms. The term from the loop quadratic in $\eta_0$ is equal to   
\begin{eqnarray}
&&\!\!\!\!\!\!\!
\frac{(3\epsilon+2\delta)^2}{256\epsilon^2}\frac{H^4}{M^4}\frac{1}{p^3}p^2\eta_0^2
\nonumber \\
&&\times
\bigg\{\Big[(1-p^2\eta^2)\cos2p(\eta-\eta_0)+2p\eta \sin2p(\eta-\eta_0)\Big]\Big[J_1-2J_0-4J_2-\frac{32}{(3\epsilon+2\delta)^2}c_3\Big] \nonumber\\
&&\quad
-(1+p^2\eta^2)J_0 \bigg\}\label{fin2}\;,
\end{eqnarray}
where
\begin{eqnarray*}
J_0 & =& \frac{1}{p^3}\int\frac{d^3\vec{q}}{(2\pi)^3}\;\frac{(p^2+q^2+k^2)^2}{qk(p+q+k)^2}\;,\\\\
J_1 &=&  \frac{1}{p^4}\int\frac{d^3\vec{q}}{(2\pi)^3}\;\frac{(p^2+q^2+k^2)^2}{qk(q+k-p)}\;,\\\\
J_2 &=& \frac{1}{p^3}\int\frac{d^3\vec{q}}{(2\pi)^3}\;\frac{(p^2+q^2+k^2)^2}{qk(p+q+k)(q+k-p)}\;.
\end{eqnarray*}
To remove it we add a quadratic term to the initial action 
\begin{eqnarray*}
S_0^{(2)} &=& \frac{1}{2}\int d^3\vec{x}\;d^3\vec{y}\;\Bigl\{\zeta^+(t_0,\vec{x})A(\vec{x}-\vec{y})\zeta^+(t_0,\vec{y})-\zeta^-(t_0,\vec{x})A^*(\vec{x}-\vec{y})\zeta^-(t_0,\vec{y})
\\
&&\qquad\qquad\qquad
+2i\zeta^+(t_0,\vec{x})B(\vec{x}-\vec{y})\zeta^-(t_0,\vec{y})\Bigr\} 
\\ 
&=& 
\frac{1}{2}\int d^3\vec{x}\;d^3\vec{y}\;\Bigl\{{\rm Re} A(\vec{x}-\vec{y})\big[\zeta^+(t_0,\vec{x})\zeta^+(t_0,\vec{y})-\zeta^-(t_0,\vec{x})\zeta^-(t_0,\vec{y})\big]
\\
&&\qquad\qquad\qquad
+{\rm Im}A(\vec{x}-\vec{y})\big[\zeta^+(t_0,\vec{x})\zeta^+(t_0,\vec{y})+\zeta^-(t_0,\vec{x})\zeta^-(t_0,\vec{y})\big]
\\
&&\qquad\qquad\qquad
+2i\zeta^+(t_0,\vec{x})B(\vec{x}-\vec{y})\zeta^-(t_0,\vec{y})\Bigr\}\;.
\end{eqnarray*}
To first order the contribution to the two-point function coming from this term is
\begin{equation}
i\langle 0|T(\zeta^+(t,\vec{x})\zeta^+(t,\vec{y})S_0^{(2)})|0 \rangle\label{fin3}\;.
\end{equation}
The part of \eqref{fin3} leading in $\eta_0$ is equal to
\begin{eqnarray}
\langle S_0^{(2)}\rangle & = &-\frac{1}{8\epsilon^2}\frac{H^4}{M^4}\frac{1}{p^6}p^2\eta_0^2 \Big\{\big[(1-p^2\eta^2)\sin2p(\eta-\eta_0)-2p\eta\cos2p(\eta-\eta_0)\big]{\rm Re}A_p\nonumber\\
&&\qquad\quad\quad\quad\quad\quad- \big[(1-p^2\eta^2)\cos2p(\eta-\eta_0)+2p\eta\sin2p(\eta-\eta_0)\big]{\rm Im}A_p\nonumber\\&&\qquad\quad\quad\quad\quad\quad+(1+p^2\eta^2)B_p\Big\}\label{fin4}\;.
\end{eqnarray}
Comparing \eqref{fin4} to \eqref{fin2} we can conclude that for $S_0^{(2)}$ to cancel the quadratically divergent terms we need

\begin{eqnarray*}
{\rm Re}A_p &=& 0\;,\\
{\rm Im}A_p & = &p^3\Big[ c_3 - {(3\epsilon+2\delta)^2\over 32} \big[\hbox{infinite part of $(J_1-2J_0-4J_2)$}\big]\Big]\\ &=& {(3\epsilon+2\delta)^2\over 32}p^3\Big[ \hbox{infinite part of $(I_4-J_1+2J_0+4J_2)$}\Big]\;,\\
B_p &=& -\frac{(3\epsilon+2\delta)^2}{16} \Big[\hbox{infinite part of $J_0$}\Big]\;.
\end{eqnarray*}

To fully renormalize the one-loop correction to the two-point function we also need to extract and eliminate from \eqref{fin1} the terms that are zeroth and first order in $\eta_0$. Since there is no principal difference between treating these terms and treating the quadratically divergent term, these further calculations are not essential for demonstrating the technique that we are introducing in this paper.    

\section{Conclusions}
For the reasons that we talked about in the introduction, it is important to be able to start the evolution of the system from a finite initial time. In this paper we presented a formalism that allows us to calculate correlation functions for states that are defined at some initial time. Using this formalism we can choose a particular state of the interacting theory at an arbitrary time, and not only in the infinite past.

 We demonstrated this technique of renormalizing the initial state for the case of the vacuum state of a toy model derived from the standard inflationary theory with a single scalar field. Using the eigenbasis of the free theory and applying matching conditions for the two- and three-point functions we were able to start constructing the initial density matrix order by order in perturbation theory: inclusion of this density matrix eliminated the unwanted finite oscillatory and divergent terms from the two- and three-point functions. 

In principle, this method can be used to renormalize other, more complicated, states, although that task might be more challenging. The main difficulty is to determine the conditions that the state should satisfy. We need to be able to translate our ideas about the physical properties of a certain state into conditions on some of its $n$-point functions. For any non-vacuum state we must start with an initial density matrix that already has some nontrivial structures. If the state we want to consider is such that it has a corresponding state in the free theory, we can start with an initial action that is only quadratic in the fields; otherwise the initial action needs to have structures of higher orders. Since we work in the free theory eigenbasis, the operators in the initial action will be defined with respect to the free theory vacuum. After applying the appropriate conditions these operators will need to be modified.

\acknowledgments

R.~H.~and T.~V.~are grateful for the support of the Department of Energy (DE-FG03-91-ER40682) and for a grant from the John Templeton Foundation.

\end{document}